\documentstyle[prd,aps,preprint,tighten]{revtex}
\begin{document}
\preprint{                                                 IASSNS-AST 96/43}
\draft
\title{         Neutrino oscillations and moments of electron spectra}
\author{                J.~N.~Bahcall and P.~I.~Krastev                    }
\address{  Institute for Advanced Study, Princeton, New Jersey 08540       }
\author{                             E.~Lisi                               }
\address{  Institute for Advanced Study, Princeton, New Jersey 08540      \\
  and Dipartimento di Fisica and Sezione INFN di Bari, 70126 Bari, Italy   }
\maketitle
\begin{abstract}
	We show that the effects of neutrino oscillations on $^8$B 
	solar neutrinos are described well by the first two moments 
	(the average and the variance) of the energy distribution of 
	scattered or recoil electrons. For the SuperKamiokande and 
	the Sudbury Neutrino Observatory experiments, the differences 
	between the moments calculated with oscillations and the 
	standard, no-oscillation moments are greater than 3 standard 
	deviations for a significant fraction of the neutrino 
	mass-mixing $(\Delta m^2,\,\sin^2 2\theta)$ parameter space. 
\end{abstract}


\section{INTRODUCTION}
\label{sec:intro}

	A full Monte~Carlo simulation is necessary in order to understand 
in detail the results of complicated experiments and to estimate their  
uncertainties. Extensive simulations will be especially important for the 
next generation of solar neutrino experiments \cite{newx} in which 
background effects, energy-dependent sensitivities, and crucial geometrical 
factors will influence the measured rates. Experimentalists will have the 
time and patience to test a finite set of hypotheses against the results 
of their measurements and massive simulations and will report the results 
to an eagerly-waiting community of physicists and astronomers.

	What additional information will be most useful to report? 

	We focus here on the crucial tests of the {\em shape\/} of the 
continuum neutrino spectrum, which are independent of solar physics to a 
fractional accuracy of ${\cal O}(10^{-5})$ \cite{shape}. Distortions of 
the energy spectra of neutrino-induced electrons could indicate neutrino 
oscillations.
  
	We show in this paper that the first two moments of the (recoil or 
scattering) electron energy distribution---the average energy and the variance
of the spectrum---can provide a compact and informative summary of the 
expected effects of neutrino oscillations on the continuum solar neutrino 
spectra. It is more practical to report the first two moments of the electron 
energy distribution and their uncertainties than to describe in detail the 
output of an extensive Monte~Carlo simulation. Moreover, theorists can 
conveniently use the measured values of the average kinetic energy and its 
dispersion, instead of a binned energy spectrum, to test different models of 
neutrino interactions.

	We calculate the changes in the first two spectrum moments, that are 
caused by  vacuum neutrino oscillations and by matter-enhanced oscillations, 
in the two-flavor approximation. We illustrate our results by realistic 
numerical studies of the SuperKamiokande \cite{SKam} (electron-neutrino 
scattering) and the Sudbury Neutrino Observatory \cite{Sudb} (SNO, neutrino 
absorption) detectors.

	The structure of our paper is the following. In Sec.~II we discuss 
the basic characteristics of the SuperKamiokande and SNO experiments, and 
present the standard (i.e., no oscillation) expectations for the average 
kinetic energy, $\langle T \rangle$, and the variance of the energy 
distribution, $\sigma^2$. In Sec.~III we describe our calculation of the
spectral moments in the presence of neutrino oscillations. In Sec.~IV we 
show the results in the neutrino mass-mixing  plane. In Sec.~V we estimate 
realistic uncertainties. In Sec.~VI we show how SuperKamiokande and SNO can 
discriminate new physics in representative cases. We summarize our work in 
Sec.~VII.

	We recall that the SNO experiment will, in addition to the electron 
spectrum shape meaurement, determine the ratio of the  charged-to-neutral 
current interactions of neutrinos in deuterium, which is a powerful 
indicator of neutrino oscillations \cite{Chen}. This measurement has been 
investigated in several  papers (see the recent works \cite{Fi94,Kr95,Li96})
and is not considered here.

\section{Basics and standard predictions}

	In this section we discuss the basic characteristics of the 
SNO and SuperKamiokande experiments, and calculate the standard electron 
energy spectra%
\footnote{	The standard electron energy spectrum at SNO has also
		been  discussed in detail in \protect\cite{Li96}.}
expected in the absence of oscillations. We also evaluate the standard
values of the first two moments of the electron energy spectra for SNO and 
SuperKamiokande. We discuss how well the two-moment approximation can 
parametrize the deviations from the standard electron spectra.

\subsection{Experimental characteristics}

	The SNO experiment \cite{Sudb} makes use of a 1 kton heavy-water 
Cherenkov detector \cite{SNOw} to observe the charged current (CC) 
neutrino-deuterium reaction:
\begin{equation}
\label{eq:nud}
\nu_e + d \to p + p + e^-\ .
\end{equation}

	The electron kinetic energy, $T$, is distributed between 0 and 
$E_\nu-Q$, where $E_\nu$ is the neutrino energy and $Q=1.442$ MeV. The 
differential cross section for this reaction has been recently discussed in 
detail in \cite{Li96}, to which the reader is referred for specific results 
and additional references \cite{cross}.

	The SuperKamiokande experiment \cite{SKam} makes use of a 22 kton 
(fiducial volume) water-Cherenkov detector \cite{SKAw} to observe 
neutrino-electron scattering. The basic reaction is:
\begin{equation}
\label{eq:nue}
\nu_e + e^- \to \nu'_e + e^-\ ,
\end{equation}
in which both charged currents (CC) and neutral currents (NC) contribute.
The electron kinetic energy $T$  is distributed almost  uniformly between 
0 and  $E_\nu/(1+2m_e/E_\nu)$. The cross-section is known  to first order 
in the radiative corrections \cite{Sirlin,cross}.

	If neutrino oscillations $\nu_e\leftrightarrow\nu_x$ to an active
neutrino occur, the pure NC reaction $\nu_x$-$e$  also contributes to the 
SuperKamiokande signal: 
\begin{equation}
\label{eq:nux}
\nu_x + e^- \to \nu'_x + e^-\ ,\quad (x=\mu,\,\tau)
\end{equation}
In the following, we assume $x=\mu$ for definiteness.

	If the SuperKamiokande and SNO detectors work as expected,  high 
signal-to-noise ratios will be achieved above a threshold energy 
$T_{\rm min}\simeq5$ MeV. The background assessment is one of the most 
difficult aspects of these experiments. In the absence of reliable 
knowledge of the background, we make the optimistic assumption that there 
is no significant background contamination above a nominal  5~MeV threshold 
for both experiments. We also assume that the detection efficiency is 
constant above threshold.

	The electron kinetic energy (and direction) will be measured, in 
both experiments, by the Cherenkov  light. The distribution
of the measured kinetic energy, $T$, around the {\em true\/} kinetic 
energy, $T'$, can be described by an energy resolution function of the form:
\begin{equation}
\label{eq:resol}
R(T,\,T') = \frac{1}{\Delta_{T'}\sqrt{2\pi}}
\exp{\left[-\frac{(T'-T+\delta)^2}{2\Delta_{T'}^2}\right]}\ .
\end{equation}

	In Eq.~(\ref{eq:resol}), the bias term $\delta$ accounts for a 
possible  uncertainty in the absolute energy calibration, and the
energy-dependent one-sigma width $\Delta_{T'}$ scales as $\sqrt{T'}$ due 
to the photon statistics:
\begin{equation}
\label{eq:DeltaT}
\Delta_{T'} =
\Delta_{10}\sqrt{\frac{T'}{10 {\rm\ MeV}}}\ ,
\end{equation}
where $\Delta_{10}$ is the energy resolution width at 10~MeV.

	The reference parameters $T_{\rm min}$, $\delta$, and $\Delta_{10}$ 
used in this work for SuperKamiokande \cite{Suzuki} and SNO \cite{Beier}
are given in Table~\ref{tab:param}. In SuperKamiokande, the energy resolution 
width $\Delta_{10}$ is expected to be measured with 1\% accuracy by means of 
a dedicated, high statistics calibration with a tunable 
linear electron accelerator \cite{Suzuki}. 

\subsection{Standard expectations}

	We have calculated the standard  (i.e, no oscillation) electron 
spectra in SuperKamiokande and SNO,  using the  detector parameters given in
Table~\ref{tab:param}, the best-estimate cross-sections for the reactions 
in Eq.~(\ref{eq:nud}) \cite{KN,Li96} and  Eq.~(\ref{eq:nue}) \cite{Sirlin}, 
and the standard $^8$B neutrino spectrum \cite{BaLi}.

	Table~\ref{tab:standard} gives the calculated values of the first 
two spectral moments of the standard distributions,  
the average electron kinetic $\langle T\rangle_0$ and the variance 
$\sigma^2_0$.

	Figure~1 compares the normalized standard spectrum (Area~=~1, solid
line labeled STD) with three  representative spectra calculated assuming 
neutrino oscillations in vacuum \cite{Pont} or oscillations enhanced 
by the Mikheyev-Smirnov-Wolfenstein (MSW) effect \cite{Wo78} in the sun.%
\footnote{	The MSW regeneration of $\nu_e$ 
		in the earth is not included in the present work.}
The  spectra with oscillations refer to the values of the neutrino 
mass-mixing parameters $\Delta m^2$ and $\sin^2 2\theta$ that best-fit 
\cite{Kr95} the  small-angle MSW (SMA), the large-angle MSW (LMA), and
the vacuum oscillation (VAC) solutions%
\footnote{	The  best-fit values $(\Delta m^2,\,\sin^2 2\theta)$
		for the SMA, LMA, and VAC cases are respectively: 
		$(5.4\times 10^{-6}{\rm\ eV}^2,\, 7.9 \times 10^{-3})$,
		$(1.7\times 10^{-5}{\rm\ eV}^2,\, 0.69)$, and
		$(6.0\times 10^{-11}{\rm\ eV}^2,\, 0.96)$ 
		\protect\cite{Kr95}.}
to the solar neutrino problem, given the published results of the four 
pioneering solar neutrino experiments \cite{Da94,GALL,SAGE,Kami}.

	The LMA solution  does not induce any appreciable difference from 
the standard spectrum, since the corresponding electron survival probability,
$P(E_\nu)=P(\nu_e\to\nu_e\,|\,E_\nu)$, is almost constant for
$E_\nu\gtrsim5$ MeV. Hence the LMA spectrum is almost indistinguishable 
from the standard spectrum in Figs.~1(a),(b). However, for the SMA and 
especially for the VAC cases illustrated, there are significant spectral 
distortions with respect to the no-oscillation (STD) case. The  deviations 
in the spectral moments, $\langle T \rangle - \langle T \rangle_0$ and 
$\sigma^2-\sigma^2_0$, are discussed in Sec.~IV.

\subsection{Two-moment approximation}

In this section we discuss the extent to which the deviations
of the first two moments characterize the deformations of the
electron spectrum induced by neutrino oscillations.

For the moderately small changes in the standard electron spectrum 
 expected on the basis of many oscillation scenarios,
the difference between the spectrum with and without oscillations
($f_{\rm OSC}(T)$ and 
$f(T)$ respectively) can be approximated by a 
second order expansion in $T$ to a good approximation.
Defining $\xi(T)$ as the difference between the spectrum after
oscillations and the standard spectrum, 
\begin{equation}
\xi(T) \equiv f_{\rm OSC}(T)-f(T)\ ,
\end{equation}
we can write
\begin{equation}
\label{eq:2order}
\xi(T) \simeq \bar\xi(T)\equiv f(T)\left(
\beta\,\frac{T-\langle T\rangle_0}{\langle T\rangle_0}+\gamma\,
\frac{T^2-\langle T\rangle_0^2-\sigma^2_0}{\langle T\rangle^2_0}\right)\ ,
\end{equation}
where $\bar\xi(T)$ is the best quadratic approximation to $\xi(T)$.

In Eq.~(\ref{eq:2order}), $\beta$ and $\gamma$ are dimensionless parameters, 
and $\langle T\rangle_0$ and $\sigma^2_0$ are the first two moments (average 
and variance) of the standard distribution $f(T)$. The functional form 
adopted in Eq.~(\ref{eq:2order}) ensures that 
$\int\! dT \bar\xi(T)=0$ and therefore preserves the normalization.

Within the ${\cal O}(T^2)$ approximation, the spectral deformations 
represent a two-parameter family of 
functions, that can be labeled either by the two parameters 
$(\beta,\,\gamma)$ or by the {\em deviations\/} 
and  of the  first two moments of 
the new distribution from their standard values, $\langle T\rangle_0$ and 
$\sigma^2_0$, through the correspondence:
\begin{eqnarray}
\langle T\rangle -\langle T\rangle_0
&\simeq&  \int\! dT\, f(T)\,\bar\xi(T;\,\beta,\,\gamma)\,T\ ,\\
\sigma^2 -\sigma^2_0
&\simeq&  \int\! dT\, f(T)\,\bar\xi(T;\,\beta,\,\gamma)\,T^2\ .
\end{eqnarray}

It is preferable to determine the deviations of the
spectral moments instead of fitting
the $(\beta,\,\gamma)$ parameters, because 
the moments are well-defined  independently of the functional form of the 
shape deformation and of the order of its expansion in powers of $T$.
 Additional experimental information 
(if any) on the third or higher moments could in principle overconstrain this 
determination, but is not essential in the two-parameter approximation of
Eq.~(\ref{eq:2order}).

	The MSW small mixing angle solution (SMA) represents an  
an important case in which  two-moment approximation
is an excellent approximation. We have checked numerically that the 
deviation $\xi(T)= f_{\rm SMA}(T)-f(T)$
between the the SMA spectrum $ f_{\rm SMA}(T)$ and the
STD spectrum $f(T)$ (see Fig.~1) can be parametrized accurately 
as indicated in Eq.~(\ref{eq:2order}). In the entire 
SMA region allowed at 95\% C.L.\ \cite{Kr95} by the four pioneering solar
neutrino experiments, the residuals, divided by the maximum spectral value,
are small:
\begin{equation}
\frac{|\xi(T)-\bar\xi(T)|}{\max f(T)}\lesssim
\left\{ \begin{array}{ll} 
1\% & \text{(Superkamiokande)}\ ,\\
3\% & \text{(SNO)}\ .
\end{array}
\right.
\end{equation}

	The large mixing angle (LMA) MSW solution is well fit by the quadratic
approximation since, in any event, the LMA spectral distortions are very small
(see Fig.~1).

	For the best-fit vacuum oscillation solution (VAC), 
the residual terms of ${\cal O}(T^3)$ or higher in the shape
deviations are as large as 10\% (5\%) of the maximum spectral value
for SuperKamiokande (SNO). The magnitude of the residuals can be made smaller,
or larger, by varying the VAC solution within the  
 region allowed at $95\%$ C.L.\ by the available data \cite{Kr95}. 
When the residuals are large, the two-moment approximation may result 
in a loss of information. 
Fortunately,  precision is not crucial in the cases in which
the distortions, $\xi(T)$, are large. In those cases, neutrino oscillations
provide a distinctive signature even with an error of
the order of 10\% in the large distortions.

	The spectral measurements
in the SuperKamiokande and SNO experiment may initially  involve 
relatively large uncertainties.
 These errors will decrease as
more events are accumulated and as the systematic uncertainties are better
understood.  If the known uncertainties are not much smaller than 
the changes in the spectrum, even a first-order approximation 
[$\gamma\simeq0$ in  Eq.~(\ref{eq:2order})] can be adequate to analyze
the data.  The spectral deformations are then in one-to-one correspondence 
either  with $\beta$ (see \cite{Kw95}), or with the deviation of 
$\langle T\rangle$  from the standard value $\langle T\rangle_0$ (deviation 
which is proportional  to $\beta$, see \cite{Li96}). The use of variations 
of $\langle T \rangle$ to signal spectral deformations was first 
emphasized 
in \cite{Fi94}(see also ref. [24]).

	In summary, we see that the description of
the spectral deformations in terms of two parameters (i.e., the 
deviations of the first two moments)
is a useful approach to the analysis of the SuperKamiokande
and SNO measurements of the electron energy distribution. The loss of
information is small for the SMA and LMA solutions. For the 
vacuum oscillation cases in which the distortion is not obvious, the first
two moments will also be an adequate description.

	We shall show below how the neutrino oscillation parameter
space can be explored by  using the deviations of the first two moments
from their standard values.

\section{Spectral moments with neutrino oscillations}

	The calculation of the  electron spectra for dense grids of values 
of the  usual neutrino mass-mixing parameters, 
$\Delta m^2$ and $\sin^2 2\theta$,  can be computer-intensive. One has to 
calculate  the differential neutrino cross-section, $d\sigma/dT$, smear it
with the energy resolution function, and integrate the results over the
$^8$B neutrino spectrum times the $\nu_e$ survival probability $P$.
However, if one is interested  in just the (first two) spectral moments, 
only one numerical integration over the neutrino energy is needed, with  
weight functions that are calculated once and for all. In this Section we 
describe this fast method of  calculation for the SNO and SuperKamiokande
experiments. The application to vacuum and matter-enhanced neutrino
oscillations will be illustrated in Sec.~IV.

\subsection{Moments of the SNO electron spectrum}

	Consider a neutrino of energy $E_\nu$ that is absorbed by deuterium,
reaction~(\ref{eq:nud}). The cross section $\hat\sigma_{\rm CC}(E_\nu)$  for 
producing an electron that has a measured kinetic energy greater than 
$T_{\rm min}$ is:
\begin{equation}
\label{eq:sigmacc}
\hat\sigma_{\rm CC}(E_\nu)=
\int_{T_{\rm min}}\!\!\!\!\!dT\int\!dT'\,R(T,\,T')\,
\frac{d\sigma_{\rm CC}(E_\nu,\,T')}{dT'} \ .
\end{equation}
where $d\sigma/dT'$ is the differential cross-section for the reaction 
(\ref{eq:nud})  \cite{KN,Li96}, and the energy resolution function
is given  in Eq.~(\ref{eq:resol}). The ``hat''  is intended to remind 
the reader that  $\hat\sigma_{\rm CC}$ is a smeared and  reduced cross 
section, i.e.,  the electrons with {\em measured\/} energy below 
$T_{\rm min}$ are not counted.

	The average value of the measured electron kinetic energy 
above threshold is:
\begin{equation}
\label{eq:hatT}
\hat T(E_\nu) = \hat\sigma^{-1}_{\rm CC}(E_\nu)
\int_{T_{\rm min}}\!\!\!\!\!dT\, T \int\!dT'\,R(T,\,T')\,
\frac{d\sigma_{\rm CC}(E_\nu,\,T')}{dT'}\ .
\end{equation}

	Analogously, the average value of the electron kinetic energy 
squared is: 
\begin{equation}
\label{eq:hatT2}
\hat T^2(E_\nu) = \hat\sigma^{-1}_{\rm CC}(E_\nu)
\int_{T_{\rm min}}\!\!\!\!\!dT\, T^2 \int\!dT'\,R(T,\,T')\,
\frac{d\sigma_{\rm CC}(E_\nu,\,T')}{dT'}\ .
\end{equation}

	So far we have considered a neutrino of fixed energy $E_\nu$. 
If we average the values of $\hat T(E_\nu)$  and $\hat T^2(E_\nu)$ 
 over the $^8$B neutrino spectrum, $\lambda (E_\nu)$, and the $\nu_e$ 
survival probability, $P(E_\nu)$, we obtain:
\begin{equation}
\label{eq:aveT}
\langle T \rangle = \frac
{\displaystyle \int\!dE_\nu\, \lambda(E_\nu)\, P(E_\nu)\, 
\hat\sigma_{\rm CC}(E_\nu)\, \hat T(E_\nu)}
{\displaystyle \int\!dE_\nu\, \lambda(E_\nu)\, P(E_\nu)\, 
\hat\sigma_{\rm CC}(E_\nu)}\ ,
\end{equation}
\begin{equation}
\label{eq:aveT2}
\langle T^2 \rangle = \frac
{\displaystyle \int\!dE_\nu\, \lambda(E_\nu)\, P(E_\nu)\, 
\hat\sigma_{\rm CC}(E_\nu)\, \hat T^2(E_\nu)}
{\displaystyle \int\!dE_\nu\, \lambda(E_\nu)\, P(E_\nu)\, 
\hat\sigma_{\rm CC}(E_\nu)}\ .
\end{equation}

	The variance of the measured electron spectrum is defined as:
\begin{equation}
\label{eq:variance}
\sigma^2=\langle T^2 \rangle-\langle T \rangle^2
\end{equation}

	The subscript ``0'' denotes ``standard'' (i.e., no oscillation)
quantities:
\begin{equation}
\label{eq:nooscill}
\langle T \rangle_0 \equiv \langle T \rangle\Big|_{P=1}\ ,\quad
\sigma^2_0\equiv \sigma^2\Big|_{P=1}
\end{equation}

\subsection{Moments of the SuperKamiokande electron spectrum}

	The calculations of the spectral moments in SuperKamiokande are 
slightly more complicated than in SNO, since the neutrino arriving at the 
detector  can be either a $\nu_e$ or a $\nu_\mu$, interacting with cross 
sections $\sigma_e$ and $\sigma_\mu$ respectively. The probabilities of 
these two occurrences are $P$ and $1-P$ respectively, where $P$ is the 
electron neutrino survival probability.

	For a  neutrino of energy $E_\nu$, we need to calculate the average 
cross section and the average electron kinetic energy for  $\alpha=e,\,\mu$: 
\begin{equation}
\label{eq:sigmalpha}
\hat\sigma_{\rm \alpha}(E_\nu)=
\int_{T_{\rm min}}\!\!\!\!\!dT\int\!dT'\,R(T,\,T')\,
\frac{d\sigma_{\nu_\alpha,e}(E_\nu,\,T')}{dT'}\ ,
\end{equation}
\begin{equation}
\label{eq:Talpha}
\hat T_\alpha(E_\nu) = \hat\sigma^{-1}_\alpha(E_\nu)
\int_{T_{\rm min}}\!\!\!\!\!dT\,T\int\!dT'\,R(T,\,T')\,
\frac{d\sigma_{\nu_\alpha,e}(E_\nu,\,T')}{dT'}\ .
\end{equation}
The expression for $\hat T^2_\alpha(E_\nu)$ is analogous to
Eq.~(\ref{eq:Talpha}) with the  substitution 
$\hat T_\alpha(E_\nu) \to\hat T^2_\alpha(E_\nu) $.

	Averaging $T$ over the neutrino spectrum, while taking into account 
the probabilities of $\nu_e$ and $\nu_\mu$ interactions at the  detector,
we obtain:
\begin{equation}
\label{eq:Tkamioka}
\langle T \rangle = \frac
{\displaystyle \int\!dE_\nu\, \lambda\, P\, \hat\sigma_e\, \hat T_e
+ \int\!dE_\nu\, \lambda\, (1-P)\, \hat\sigma_\mu\, \hat T_\mu}
{\displaystyle \int\!dE_\nu\, \lambda\, P\, \hat\sigma_e\, 
+ \int\!dE_\nu\, \lambda\, (1-P)\, \hat\sigma_\mu}\ .
\end{equation}
For clarity, we have suppressed the $E_\nu$-dependences in 
Eq.~(\ref{eq:Tkamioka}).

	The calculation of $\langle T^2\rangle$ is analogous to 
Eq.~(\ref{eq:Tkamioka}), with the substitution 
$\hat T_\alpha\to\hat T^2_\alpha$.
The variance $\sigma^2$ is defined as in Eq.~(\ref{eq:variance}).

\section{Results in the mass-mixing plane 
$(\Delta {\lowercase{m},\,\sin^2 2\theta}^2)$}

	From the discussion of the  previous section, it follows that 
the calculation of the spectral moments in SNO and SuperKamiokande 
requires integrating over energy various products of the following 
quantities: 
$\lambda(E_\nu)$ ($^8$B neutrino spectrum);
$\hat\sigma_{\rm CC}(E_\nu)$, $\hat T (E_\nu)$, and $\hat T^2(E_\nu)$
(SNO experiment); 
$\hat\sigma_e(E_\nu)$, $\hat T_e (E_\nu)$,  
$\hat T^2_e (E_\nu)$,
$\hat\sigma_\mu(E_\nu)$, $\hat T_\mu (E_\nu)$, and  
$\hat T^2_\mu (E_\nu)$ (SuperKamiokande experiment).
These ingredients can be calculated  at representative neutrino energies,
given the detector parameters (Table~\ref{tab:param}).

	A numerical table of the above quantities can be found in
\cite{cross}. The reader can find there also tables of  the last ingredient 
needed to calculate $\langle T \rangle$ and $\sigma^2$, namely the 
oscillation probability $P(E_\nu)$, for the best-fit cases SMA, LMA, VAC 
shown in Fig.~1. Anyone wishing to evaluate the sensitivity of 
SuperKamiokande or SNO to some neutrino model not considered in this paper
(e.g., resonant magnetic transitions, neutrino decay, violation of
the equivalence principle, or exotic transitions \cite{other}) can use the
data and software in \cite{cross}.

	We have calculated the values of the spectral moments 
$\langle T \rangle$ and $\sigma^2$ for continuous ranges of the  
neutrino mass-mixing parameters for two-flavor oscillations. In particular, 
we have  considered the rectangle 
$(\sin^2 2\theta,\,\Delta m^2/{\rm eV}^2)\in
[10^{-4},\,1]\otimes [10^{-9},\,10^{-3}]$, which is relevant to MSW
oscillations, and the rectangle 
$(\sin^2 2\theta,\,\Delta m^2/{\rm eV}^2)\in
[0.4,\,1]\otimes [3\times 10^{-11},\,2\times10^{-10}]$, which is 
relevant to vacuum neutrino oscillations.

	Figure~2 shows the calculated results for the SuperKamiokande 
experiment and  Fig.~3 gives the results for the SNO experiment.

	Figures~2(a),(b) show the fractional differences $(\%)$ in the first
two moments of the SuperKamiokande electron spectrum,
$(\langle T\rangle -\langle T\rangle_0)/\langle T\rangle_0$
and $(\sigma^2-\sigma^2_0)/\sigma^2_0$, for MSW neutrino oscillations.
Superposed are the regions allowed at $95\%$ C.L.\ by current fits to solar 
neutrino data \cite{Kr95} (shaded), with thick dots marking
the best-fit points.

	The deviations of the moments are larger in two characteristic 
regions, corresponding to the  adiabatic MSW branch (horizontal region), 
and non-adiabatic MSW branch (slanted region). In the adiabatic branch, 
there is a strong suppression of the high-energy part of the $^8$B neutrino 
spectrum. Therefore the mean value of the electron energy also decreases, 
until there are very few electrons above the  experimental threshold 
energy $T_{\rm min}$ (5 MeV nominal value). The variance is also reduced 
by the combination of the depletion of the high energies by oscillations 
and the threshold cut. The  fractional spectral variations are so large 
that part of the adiabatic branch is already excluded by the 
``low statistics'' electron spectrum measurements at Kamiokande 
\cite{Kspectrum}. For the non-adiabatic branch (relevant for the
small mixing angle solution) the low-energy part of the $^8$B neutrino
spectrum is suppressed,  so that both $\langle T\rangle$ and $\sigma^2$ are
shifted towards higher values. The changes in the electron spectrum
are typically smaller\cite{Bethe} than in the adiabatic region. 
In the large mixing angle region, the suppression of the $^8$B neutrino
spectrum is almost uniform in the energy range of interest and there 
are no significant deviations of the moments from their standard values.

	Figures 2(c),(d) show the fractional differences in the first
two moments of the SuperKamiokande electron spectrum for vacuum neutrino 
oscillations. Also in this case there are two regions where the moments 
are positively or negatively shifted, corresponding  to the suppression of 
low-energy and high-energy $^8$B neutrinos. In passing from one region to 
another, there are also cases in which the two moment deviations have 
opposite sign. These intermediate cases correspond roughly to a strong 
suppression of the peak of the  $^8$B neutrino spectrum, and thus a 
secondary signature should be a  low rate of events. The variations of 
the moments can be very strong for large values of the mixing angles. 
Asymptotically, they become zero for $\Delta m^2\to\infty$ (averaged fast 
oscillations, uniform suppression), and for $\Delta m^2\to 0$ or
$\sin^2 2\theta\to 0$ (no oscillation).

	Figs.~3(a--c) are analogous to Figs.~2(a--c), but refer to the 
SNO experiment. In general, the deviations of the moments are larger than 
in SuperKamiokande, since the final  state electrons from $\nu$-$d$ 
absorption  are more closely correlated with the original neutrino 
energies than the electrons from $\nu$-$e$ scattering, and thus are 
more sensitive to oscillations of the parent neutrinos. We will recall 
this point in Sec.~VI.

	Figures analogous to Fig.~2(a) and 3(a) were first reported
in \cite{Fi94}. In \cite{Fi94}, the energy resolution function was not 
included, and the electron energy spectrum  in $\nu$-$d$ absorption processes
was approximated by a Dirac delta at any given  neutrino energy 
(which is not a sufficient approximation, see the discussion in 
\cite{Li96}). Alternative  methods for analyzing  electron spectra 
have been proposed in \cite{Kw95,Smirnov}.

\section{Error evaluation}

	In order to test the null hypothesis of no oscillations,
we need to compare the theoretical spectral moment deviations  shown in 
Figs.~2 and 3 with  realistic error estimates for $\langle T\rangle_0$  
and $\sigma^2_0$.

	Table~\ref{tab:standard} lists  the standard  values of the first
two moments,  $\langle T\rangle_0$  and $\sigma^2_0$.  We will now  
estimate the errors affecting these values by varying the input ingredients 
of our calculations. A more realistic assessment of the uncertainties  
will be possible after the SuperKamiokande and SNO experiments  
have reported on  detector performances.

	Table~\ref{tab:errors} shows a preliminary assessment of the error 
budget for SuperKamiokande and SNO.  In Table~\ref{tab:errors},  the errors 
due to the energy resolution width and energy scale  are obtained by 
varying the parameters $\Delta_{10}$ and $\delta$ within the $1\sigma$ 
limits given in Table~\ref{tab:param}. The error due to the $^8$B neutrino 
spectrum is obtained by  repeating the calculation with the $\pm3\sigma$  
$^8$B spectra, $\lambda^+$ and $\lambda^-$ \cite{BaLi}, and dividing the 
total spread  by six. The cross section uncertainty for the SNO predictions 
has been  estimated by comparing the values obtained with the Kubodera-Nozawa 
cross-section \cite{KN} (default) and with the Ellis-Bahcall \cite{Ellis,Li96}
cross-section for the $\nu$-$d$ CC reaction. For SuperKamiokande, the 
$\nu$-$e$ cross-section is known  to first order in the radiative 
corrections \cite{Sirlin}.  The first order contributions  
(included by default) alter $\langle T \rangle$ and $\sigma^2$  by $0.1\%$ 
and $0.4\%$ with respect to the tree-level cross-section \cite{Ba89}. 
The contributions due to second and  higher order corrections are expected
to be smaller by  a factor of order $1/137$ and can thus be  neglected.
The statistical errors are calculated for a representative case
of $N=5000$ collected events, which corresponds roughly to 1--2 years
of operation, depending on the absolute $^8$B neutrino flux and the
oscillation scenario. The error of  $\langle T \rangle$ is given
by $\sqrt{\sigma^2/N}$. The error of $\sigma^2$ is given by 
$\sqrt{[\mu_4-(\sigma^2)^2]/N}$ \cite{Kendall}, where $\mu_4$ is the fourth 
moment of the spectrum, $\mu_4=\langle (T-\langle T \rangle)^4 \rangle$.
As stated earlier, we have not included uncertainties due to the backgrounds;
these uncertainties may be important.

	Let us consider the error correlations in Table~\ref{tab:errors}.
The correlation between the  errors on $\langle T\rangle_0$ and $\sigma^2_0$ 
induced by the uncertainties in the energy resolution width $\Delta_{10}$
(see Table~\ref{tab:param}) is equal to one in modulus, since both errors 
depend on the same parameter ($\Delta_{10}$). The sign of the correlation 
is $+$ because any small positive shift in $\langle T \rangle$ tends to 
make the low-energy part of the electron spectrum wider (above threshold) 
and thus increases  $\sigma^2$. For similar reasons, the errors induced by 
the energy scale uncertainty, as well as by the neutrino spectrum 
uncertainties, have correlation $+1$. The cross-section errors are 
small (negligible for SuperKamiokande) and  their correlation can be 
ignored. The correlation of the statistical errors is
$\rho=\mu_3/[\sqrt{\sigma^2}\sqrt{\mu_4-(\sigma^2)^2}]$ \cite{Kendall},
where $\mu_3$ is the third moment of the electron energy spectrum.

\section{Iso-sigma contours for SuperKamiokande and SNO}

	In this Section we illustrate the diagnostic power of SuperKamiokande
and SNO to reveal possible new physics by measuring electron energy spectra.

	Figure~4 shows contours of equal standard deviations 
($n$-sigma ellipses)%
\footnote{	The ``number of sigmas'' $n$ is defined as
		$n=\protect\sqrt{\chi^2}$. The probability content of the
		error ellipses is given by the $\chi^2$ distribution for 
		two degrees of freedom \cite{PDGR}.}
in the plane of the $\langle T \rangle$ and $\sigma^2$ deviations that were 
obtained with the help of Table~\ref{tab:errors}. The contours are centered
around the standard expectations (STD). Also shown are the representative 
best-fit points VAC and SMA.  The point LMA is very close to STD and is 
not shown.

	The cross centered at the SMA best-fit point indicates the solution 
space allowed at  $95\%$ C.L.\ \cite{Kr95} by the solar neutrino data 
published so far  [see Figs.~2(a), 2(b), 3(a), and 3(b)]. The deviations 
in $\langle T \rangle$  and $\sigma^2$ for the SMA solution are confined 
in a  relatively small range. For vacuum oscillations, the range of 
deviations spanned by the whole region allowed at 95\% C.L.\ by present 
data  [Figs.~2(c), 2(d), 3(c), 3(d)] is much larger and is not indicated 
in Fig.~4. The statistical significance of the separation between the SMA 
and STD points  in Fig.~4  is dominated by the fractional shift in 
$\langle T \rangle$ for both SuperKamiokande and SNO. This is not surprising, 
since the SMA neutrino survival probability increases almost linearly 
with energy for $E_\nu\gtrsim 5$ MeV; this increase  induces deformations 
of the electron spectra that are nearly linear in $T$ and are well 
represented by a shift in $\langle T \rangle$, as discussed in Sec.~II
(see also \cite{Li96}).

	The best-fit small mixing angle solution is separated by 
$\gtrsim 3\sigma$ from the standard solution for both SuperKamiokande
and SNO (see Fig.~4). The discriminatory power of the  two experiments 
appears to be comparable for the SMA solution. The  estimated total 
fractional errors of $\langle T \rangle$ and $\sigma^2$ in SuperKamiokande 
are about a factor of two smaller than in SNO (see Table~\ref{tab:errors}). 
However, the purely CC interaction in SNO [Eq.~(\ref{eq:nud})] 
is a more sensitive probe of neutrino oscillations than a linear 
combination of CC and NC interactions, as observed
in SuperKamiokande  [Eqs.~(\ref{eq:nue}) and (\ref{eq:nux})].%
\footnote{In practice, the separation of CC and NC events in SNO will
be affected by experimental uncertainties. We have ignored 
this mis-identification error as well as all the backgrounds.}
Moreover, the final electron energy in the 
$\nu_e+d\to p+p+e^-$ absorption 
interaction  is more closely correlated with the initial neutrino energy
than in $\nu+e^-\to \nu'+e^-$ scattering. The SuperKamiokande detector
(according to the available information about the likely performances of 
the detectors) may compensate a lower sensitivity to 
neutrino oscillations 
with a more precise control of the systematics related to the
electron energy measurement. 

How do the above results depend upon the energy threshold?
We have verified by detailed calculations that 
the statistical 
significance of the SMA deviations in Figs.~4(a) and 4(b) decreases by 
about $0.6\sigma$ per 1 MeV increase in the  energy threshold 
$T_{\rm min}$.  These results are valid for both the SNO and the
SuperKamiokande detectors and include calculations for thresholds of
$5$, $6$, and $7$ MeV.

	Although  SuperKamiokande and SNO are calculated to be about 
equally sensitive to the SMA solution, different new physics scenarios may 
be more easily tested by one or the other of the two experiments. For 
instance, the SNO experiment appears to be more sensitive than 
SuperKamiokande to the best-fit vacuum solution
(VAC point in Fig.~4).

	When both the SuperKamiokande and the SNO electron spectra
have been measured, it will be important to estimate the correlation of their
errors.  In fact, the uncertanties affecting the spectra measured in these
two different experiments are  all independent of each other except for
the uncertainty in the shape of the input neutrino spectrum \cite{BaLi}, 
which is common to both experiments. The uncertainties in the neutrino 
spectrum  will induce a correlation between the errors in the spectrum 
shapes measured by SNO and SuperKamiokande. 
For the spectral moments and the detector parameters used in this work, 
the correlation matrix between the total errors of Superkamiokande and SNO 
is shown in  Table~\ref{tab:correl}. The submatrix that links SNO and 
SuperKamiokande quantities has relatively small entries, $\rho=0.12$--$0.17$, 
indicating that the ``cross-talk''  induced by the common neutrino spectrum 
systematics is relatively unimportant. However, this relatively small error 
correlation may increase or decrease for more realistic  detector 
parameters and experimental uncertainties, that will be determined after 
the experiments have been operating for some time.

\section{Summary}

	In this paper we propose that the electron energy spectra 
that will be measured in the SuperKamiokande and SNO experiments
be characterized by their first two moments, namely the average 
electron kinetic energy,  $\langle T\rangle$, and the variance of the 
energy distribution, $\sigma^2$. 

	We have shown in Sec.~II~C that the two-moment approximation is 
accurate for the MSW solutions favored by the 
four pioneering solar neutrino
experiments \cite{Da94,GALL,SAGE,Kami} and is a good approximation to
the distortions induced by vacuum oscillations except for those cases
in which the distortion is very large and obvious.

	We have presented in Tables~\ref{tab:standard} 
and \ref{tab:errors}
the standard values, and estimated uncertainties, of $\langle T\rangle$ and 
$\sigma^2$ in the absence of oscillations. We have
also presented in Figs.~2 and 3
the fractional deviations of the two spectral moments induced by neutrino 
oscillations in the whole neutrino mass-mixing parameter space relevant to  
MSW or vacuum two-flavor oscillations. The deviations from the standard
values can be greater than $3\sigma$, as shown in  Fig.~4, in a 
large region 
of the neutrino parameter space, 
part of which is currently favored by the 
published results  of the four pioneering solar neutrino experiments.
The statistical significance of the SMA deviations for both SNO and
SuperKamiokande decrease by about $0.6\sigma$ per MeV for energy
thresholds in the range between $5$ and $7$ MeV.\footnote{After this
paper was completed, H. Sobel suggested we investigate 
what would be the
effect of reducing the uncertainty in the energy scale, $\delta$, in the 
SuperKamiokande experiment
from $100$ keV to $50$ keV.  We find that if $\delta$ is reduced to 
$50$ keV
the total uncertainty in
both moments is reduced by about $15$\% (5000 events assumed) and the
significance of the SMA signal is increased by about $0.5\sigma$ (to
about
$3.5\sigma$). The uncertainty in the $^8$B neutrino spectrum would be
the dominant systematic uncertainty (excluding background effects) if
the energy scale error were as small as $50$ keV.}
\acknowledgements

	We thank Y.~Totsuka and Y.~Suzuki for reading a draft manuscript
and for useful information about
the SuperKamiokande experiment. JNB acknowledges support from NSF grant 
No.\ PHY95-13835.  The work of PIK was partially supported by funds of 
the Institute for Advanced Study. The work of EL  was supported in part by 
INFN and in part by a Hansmann membership at the Institute for Advanced 
Study, and was performed under the auspices of the European Theoretical and 
Astroparticle Network (TAN).


\begin{table}
\caption{	Electron energy threshold and resolution parameters 
		adopted in the analysis. Uncertainties are at $1\sigma$.}
\smallskip
\begin{tabular}{lccc}
Detector parameter &
Symbol                        &   SuperKamiokande   &        SNO          \\
\hline
Kinetic energy threshold      &
$T_{\rm min}$                 &        5 MeV        &       5 MeV         \\
Resolution width at 10 MeV    & 
$\Delta_{10}$                 &$1.6(1\pm0.01)$ MeV & $1.1(1\pm0.10)$ MeV  \\
Energy scale uncertainty      &
$\delta$                      &    $\pm100$ keV     &   $\pm100$ keV      \\
\end{tabular}
\label{tab:param}
\end{table}

\begin{table}
\caption{	Standard predictions for the average kinetic energy,
		$\langle T \rangle$, and for the variance, $\sigma^2$,
		of the electron spectra at SuperKamiokande and SNO in 
		the absence of oscillations.}
\smallskip
\begin{tabular}{lcc}
Experiment         & $\langle T\rangle_0$ (MeV) & $\sigma^2_0$ (MeV$^2$)\\
\hline
SuperKamiokande   &     7.296 & 3.42 \\
SNO		  &     7.658 & 3.04 
\end{tabular}
\label{tab:standard}
\end{table}

\begin{table}
\caption{	Estimated $1\sigma$ uncertainties $(\Delta)$ of the 
		standard moments $\langle T \rangle_0$ and $\sigma^2_0$, 
		and their correlations $(\rho)$. Possible background effects 
		are ignored.} 
\smallskip
\begin{tabular}{lccccccc}
&            \multicolumn{3}{c}{SuperKamiokande}&&\multicolumn{3}{c}{SNO}\\
                                 \cline{2-4}                  \cline{6-8}
&$\Delta\langle T\rangle_0$&
$\Delta\sigma^2_0$&&&$\Delta\langle T\rangle_0$&$\Delta\sigma^2_0$&  \\
$1\sigma$ uncertainties &(MeV)&(MeV$^2$)&$\rho$ &&(MeV)&(MeV$^2$)&$\rho$ \\
\hline
Energy resolution width & 0.004 & 0.02  & +1    && 0.024 & 0.10 & +1     \\
Energy scale            & 0.025 & 0.06  & +1    && 0.052 & 0.08 & +1     \\
$^8$B neutrino spectrum & 0.016 & 0.04  & +1    && 0.029 & 0.05 & +1     \\
Cross section           &$\sim0$&$\sim0$&$\sim0$&& 0.011 & 0.01 &$\sim0$ \\
Statistics ($N=5000$)   & 0.026 & 0.08  & +0.63 && 0.025 & 0.06 &+0.45   \\
\hline
Total $1\sigma$ error   & 0.040 & 0.11  & +0.80 && 0.070 & 0.15 & +0.83  \\
Total fractional error  & 0.54\%& 3.2\% & +0.80 && 0.91\%& 4.9\%& +0.83  \\
\end{tabular}
\label{tab:errors}
\end{table}

%
%
%
%

\begin{table}
\label{tab:correl}
\caption{	The correlation matrix of the total errors affecting the 
		measurements of $\langle T\rangle_0$ and $\sigma^2_0$ at 
		SuperKamiokande (SK) and SNO. The correlations between the 
		SK and SNO entries  arise from the  common uncertainty in 
		the $^8$B neutrino spectral shape.}
\smallskip
\begin{tabular}{c|cccc}
&                    $\langle T\rangle_0$\ (SK)  &$\sigma^2_0$\ (SK)        
&                    $\langle T\rangle_0$\ (SNO) &$\sigma^2_0$\ (SNO)\\
\hline
$\langle T\rangle_0$  (SK)  &  1.00  &  0.80  &  0.17  &  0.13       \\
$\sigma^2_0$\ (SK)          &        &  1.00  &  0.15  &  0.12       \\
$\langle T\rangle_0$\ (SNO) &        &        &  1.00  &  0.83       \\
$\sigma^2_0$\ (SNO)         &        &        &        &  1.00       \\
\end{tabular}
\end{table}


\begin{figure}
\caption{	Standard spectrum of  the electron kinetic energy 
		(STD, solid line) for SuperKamiokande (a) and SNO (b). 
		Also shown are three representative spectra that apply 
		if neutrino oscillations occur: small mixing angle MSW 
		(SMA, dot-dashed),  large mixing angle MSW (LMA, dashed), 
		and vacuum oscillations (VAC, dotted). All spectra are 
		normalized (Area~=~1). The oscillation cases correspond 
		to  best fits  \protect\cite{Kr95} to the results of the 
		four pioneering solar neutrino experiments 
		\protect\cite{Da94,GALL,SAGE,Kami}.}
\end{figure}

\begin{figure}
\caption{	SuperKamiokande experiment:  fractional deviations (\%) 
		of the moments of the electron energy spectrum caused by 
		two-flavor neutrino oscillations. The results are shown in 
		the mass-mixing plane $(\sin^22\theta,\,\Delta m^2)$. The 
		shaded regions are favored by current solar neutrino
		experiments at $95\%$ C.L.\ \protect\cite{Kr95}, with 
		best-fit points marked by a  dot. 
		(a) Deviation of $\langle T\rangle$, MSW oscillations.
		(b) Deviation of $\sigma^2$, MSW oscillations. 
		(c) Deviation of $\langle T\rangle$, vacuum oscillations.
		(d) Deviation of $\sigma^2$, vacuum oscillations.}
\end{figure}

\begin{figure}
\caption{	SNO experiment: fractional deviations (\%) of the moments 
		of the electron energy spectrum caused by two-flavor 
		neutrino oscillations. The results are shown in the 
		mass-mixing plane $(\sin^22\theta,\,\Delta m^2)$. The 
		shaded regions are favored by current solar neutrino
		experiments at $95\%$ C.L.\ \protect\cite{Kr95}, with 
		best-fit points marked by a  dot. 
		(a) Deviation of $\langle T\rangle$, MSW oscillations.
		(b) Deviation of $\sigma^2$, MSW oscillations.
		(c) Deviation of $\langle T\rangle$, vacuum oscillations.
		(d) Deviation of $\sigma^2$, vacuum oscillations.}
\end{figure}

\begin{figure}
\caption{	Iso-sigma contours in the plane of the fractional deviations 
		of the first two spectral moments. Labels are as in Fig.~1. 
		(a) SuperKamiokande experiment. 
		(b) SNO experiment. 
		The SMA solution can be distiguished at 
		$\protect\gtrsim3\sigma$ from the standard (STD) case
		by both experiments. The crosses allow for variations of 
		the SMA solution within the region  favored at $95\%$ C.L.\ 
		by the current experiments. See the text for details.}
\end{figure}

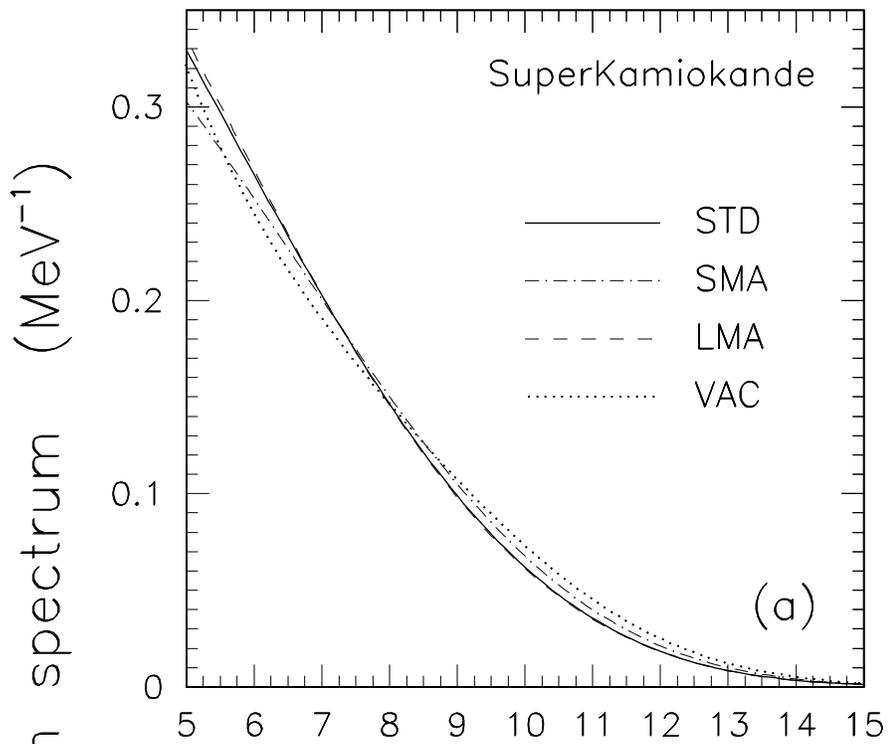
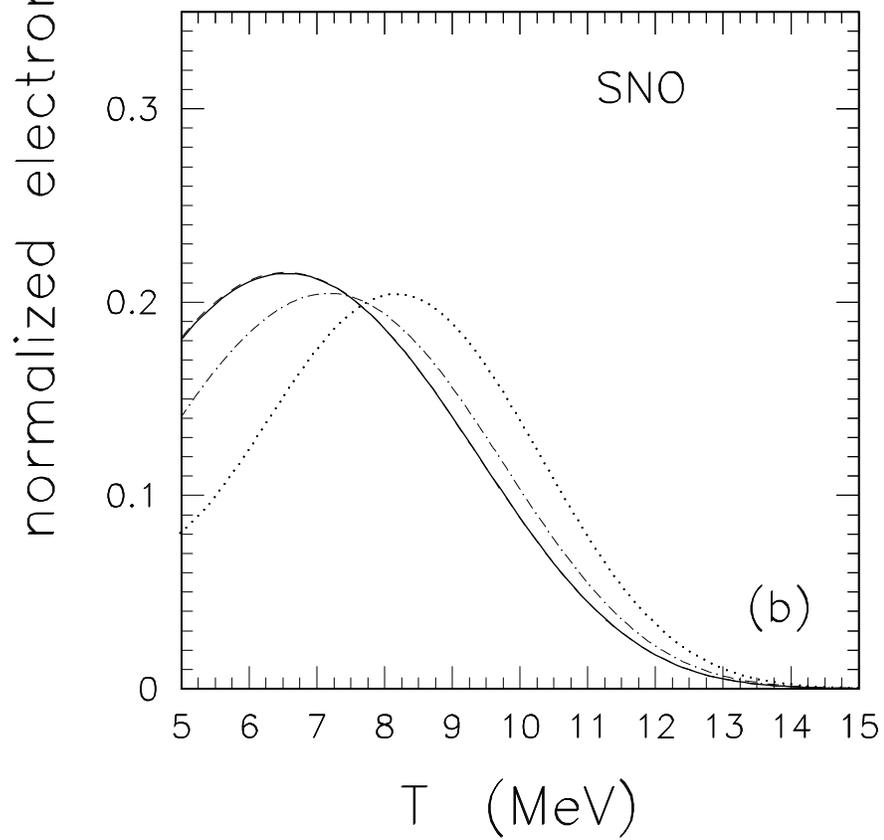
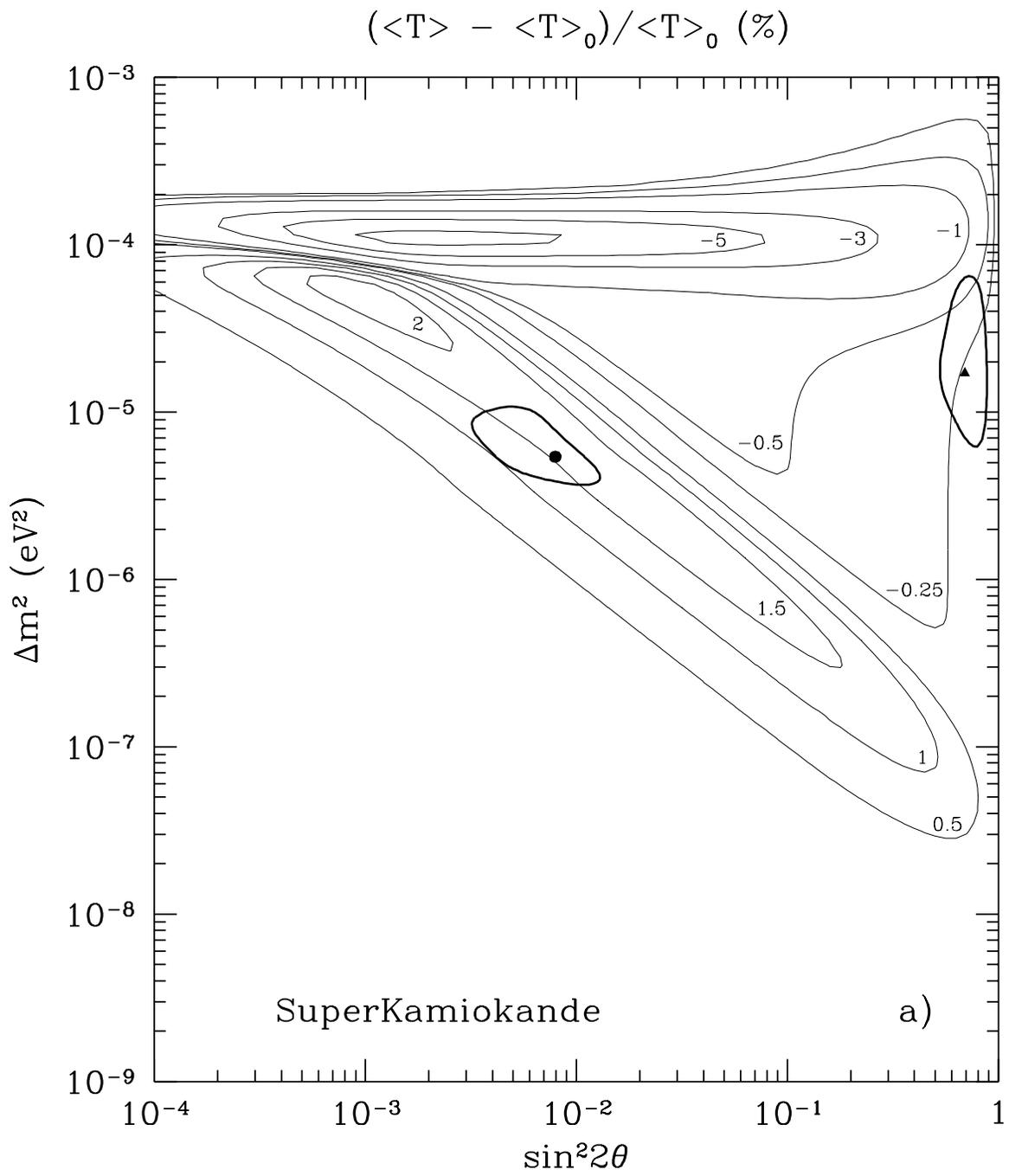
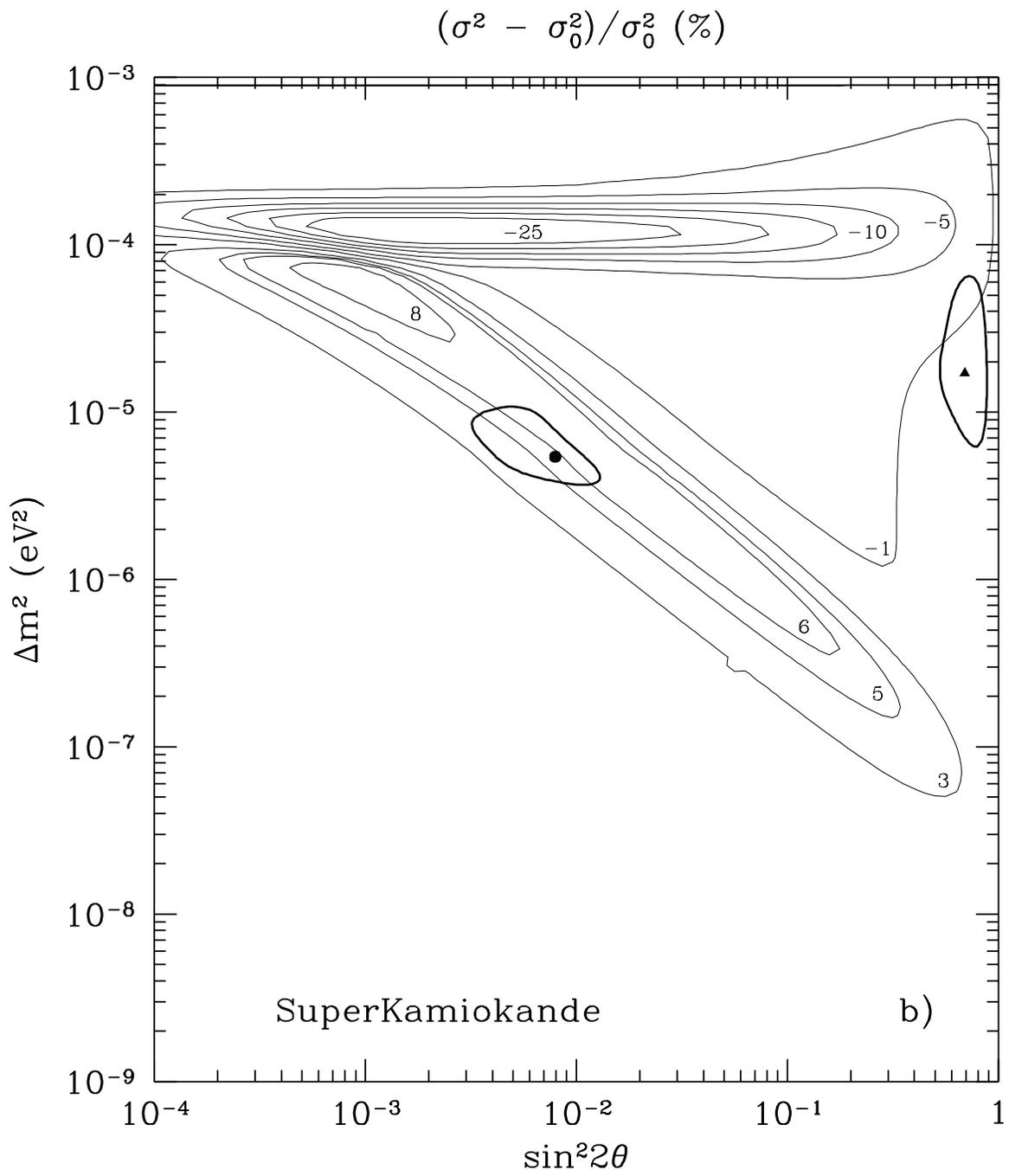
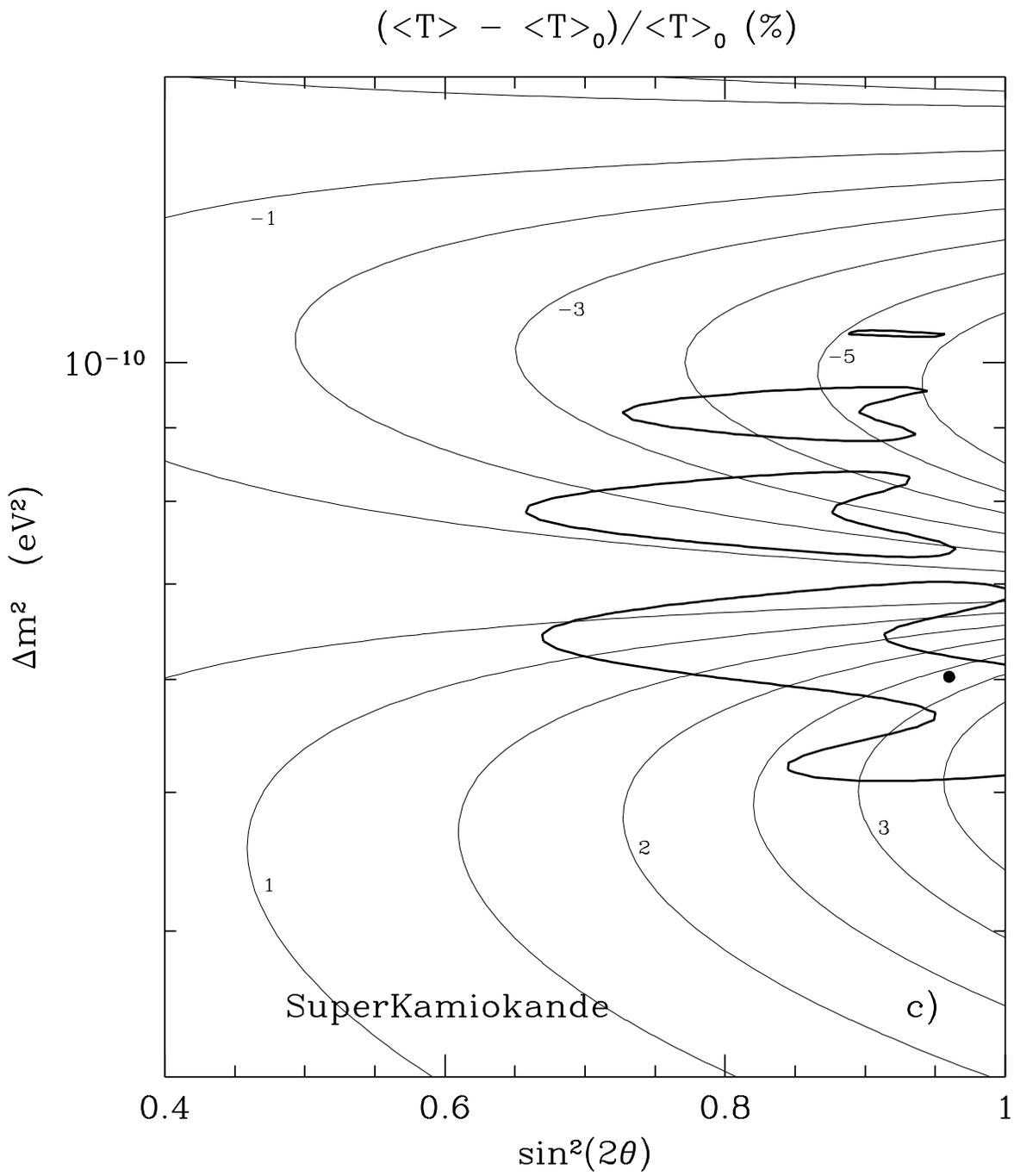
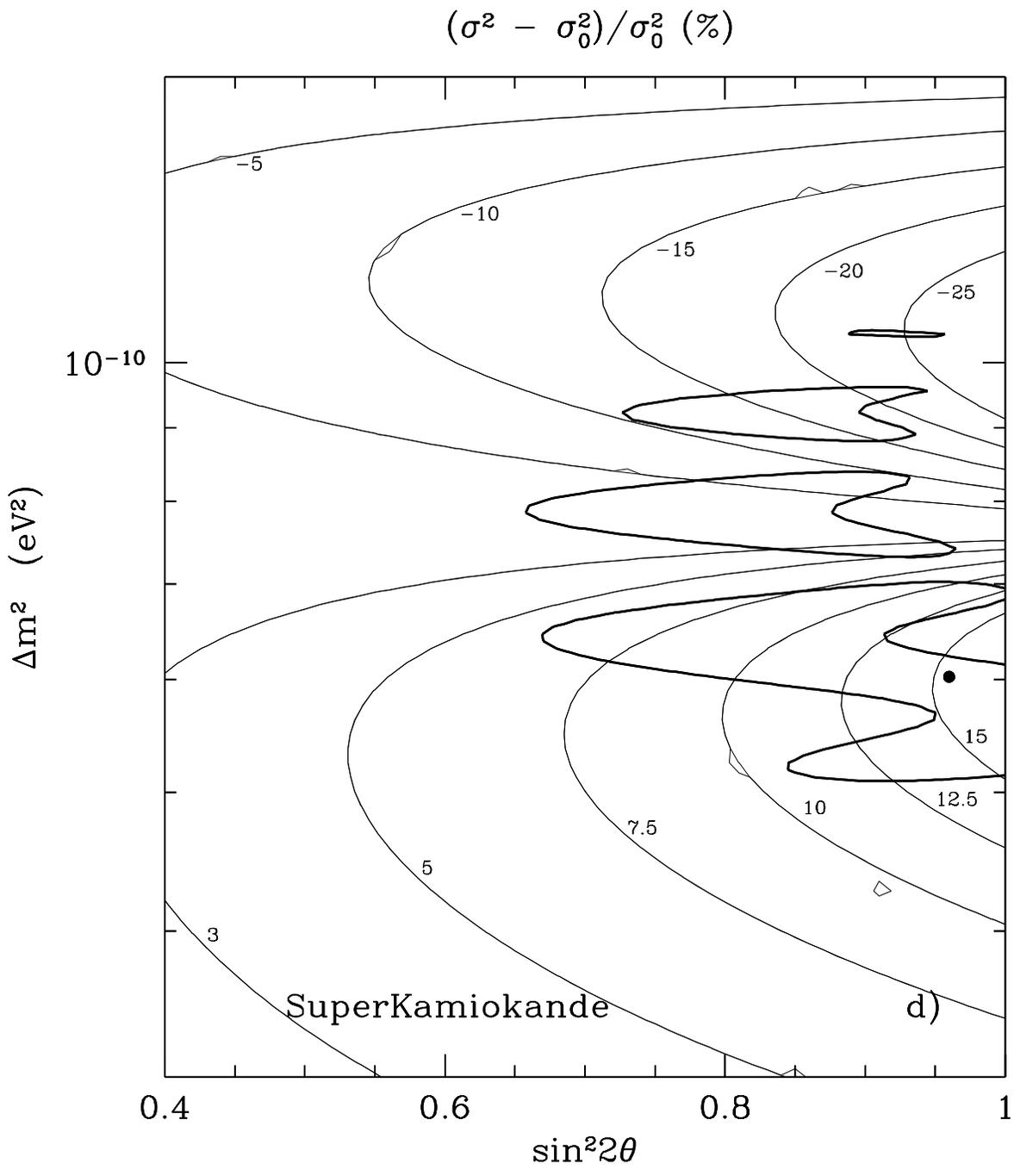
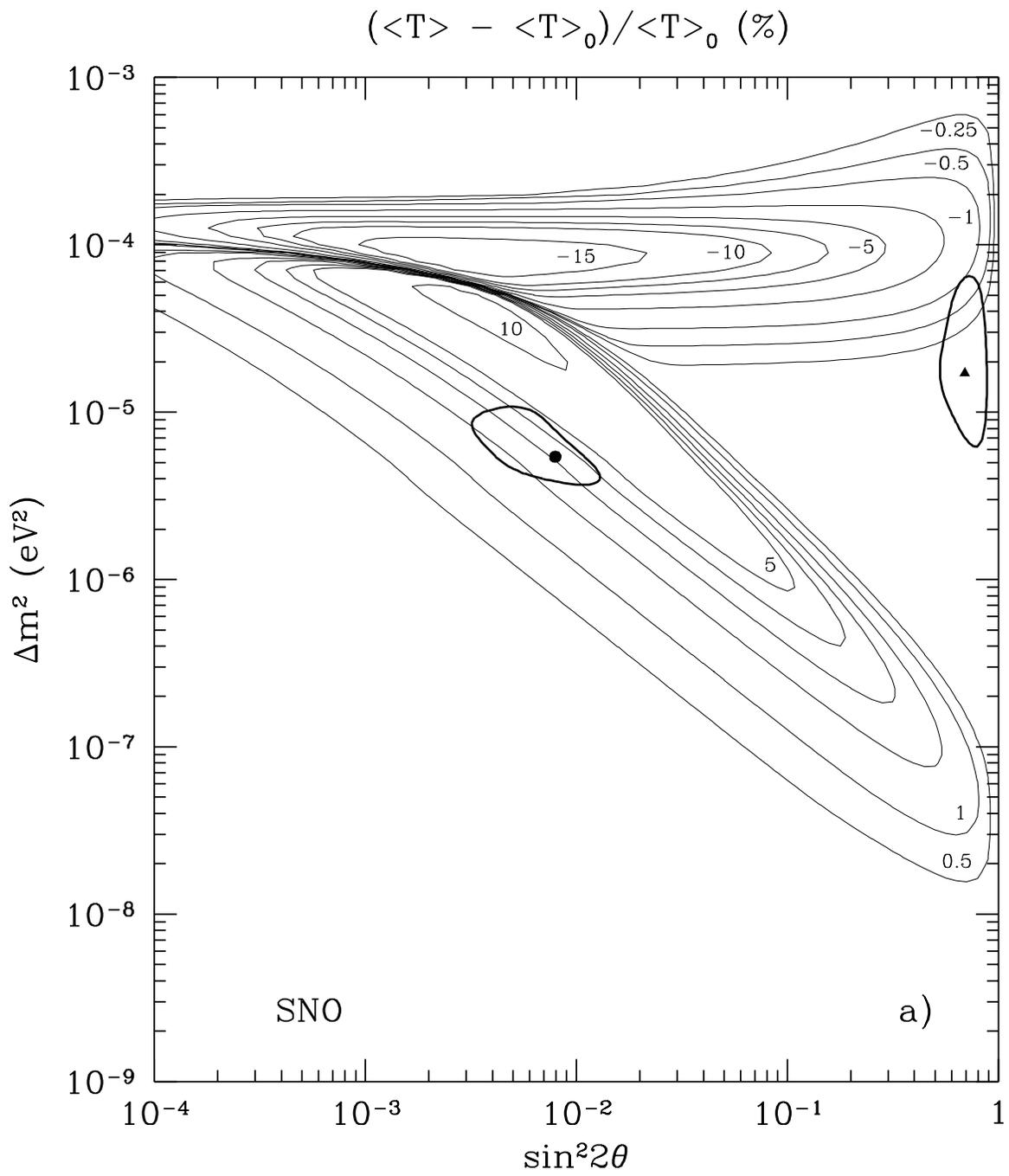
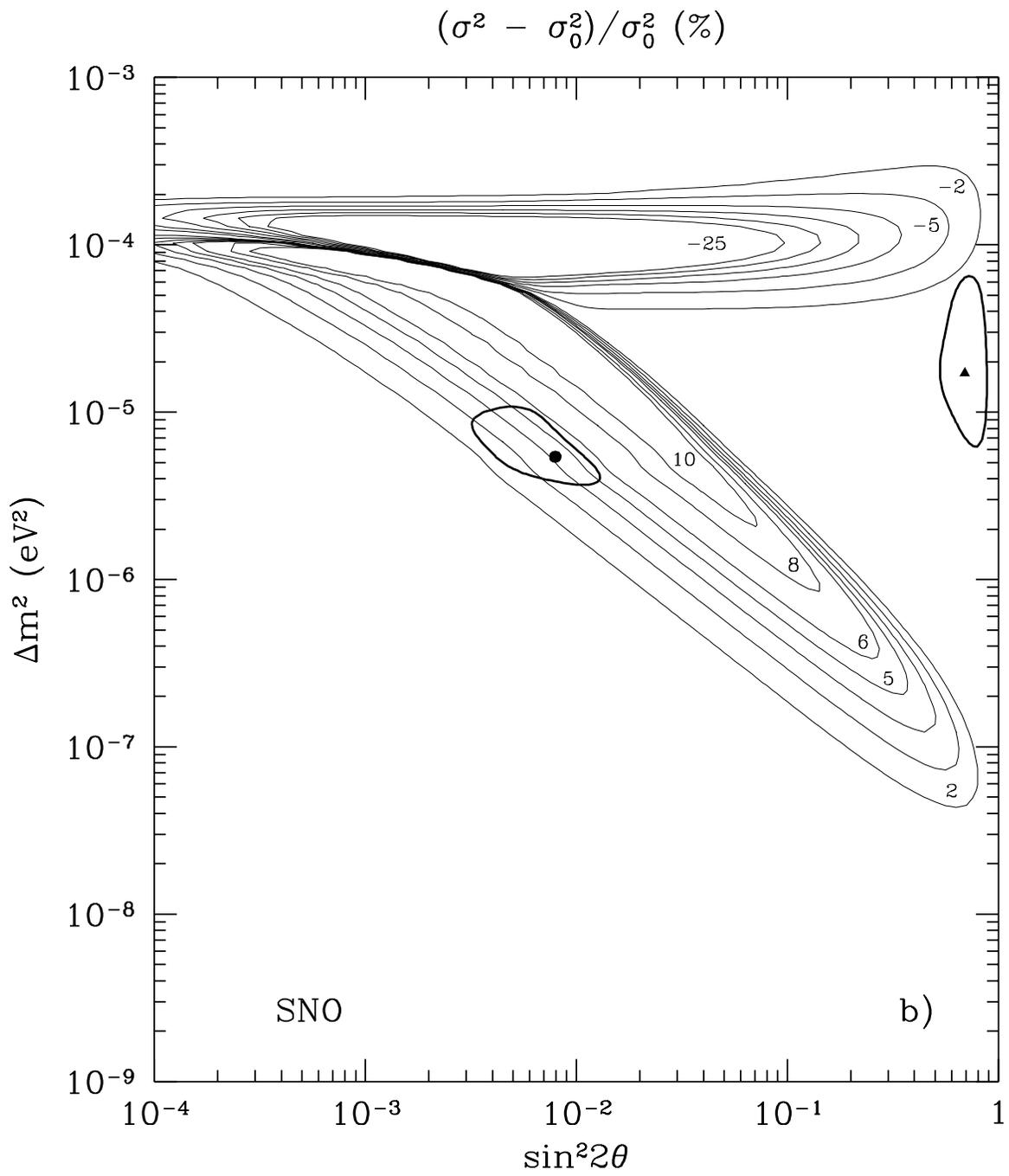
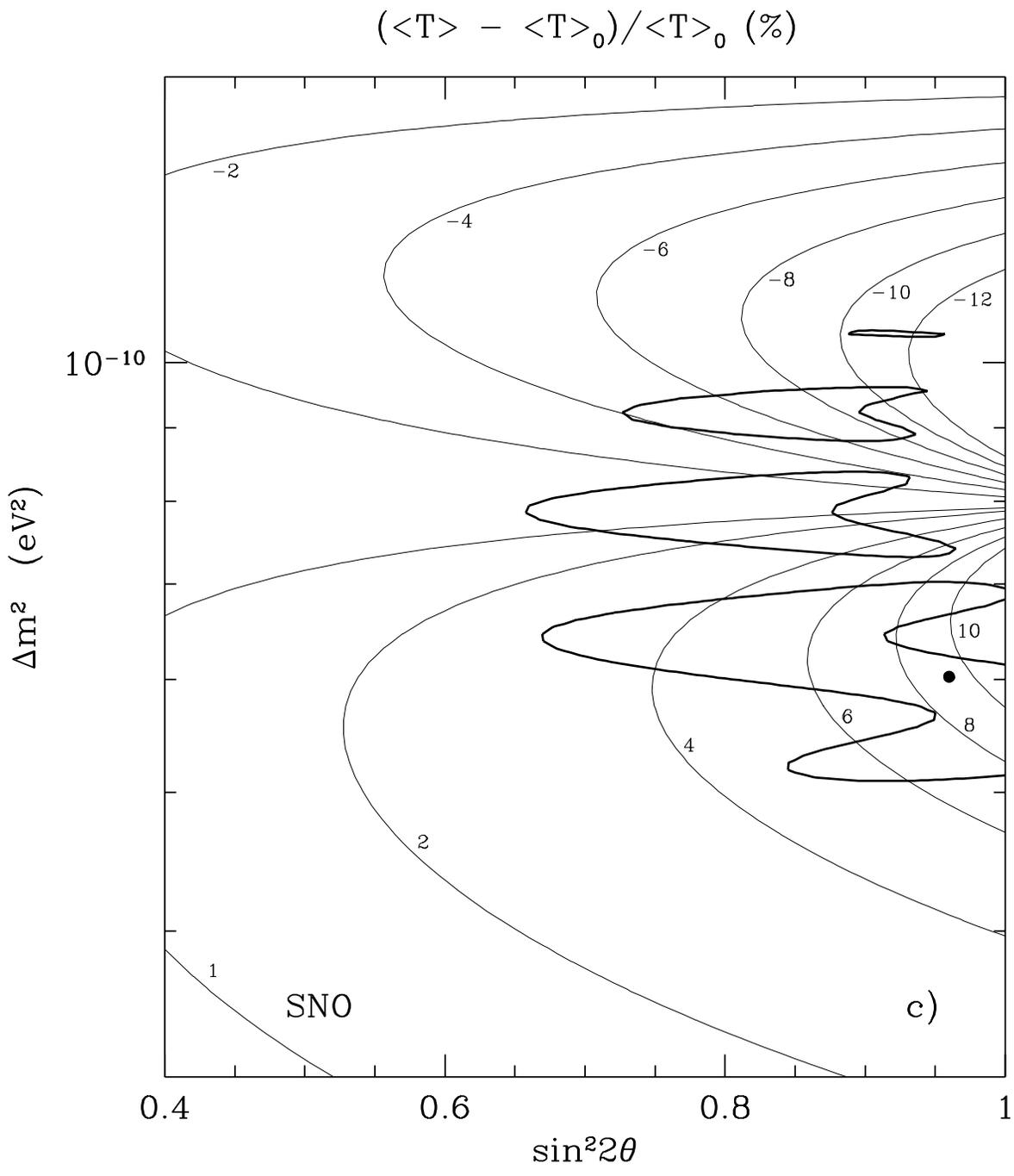
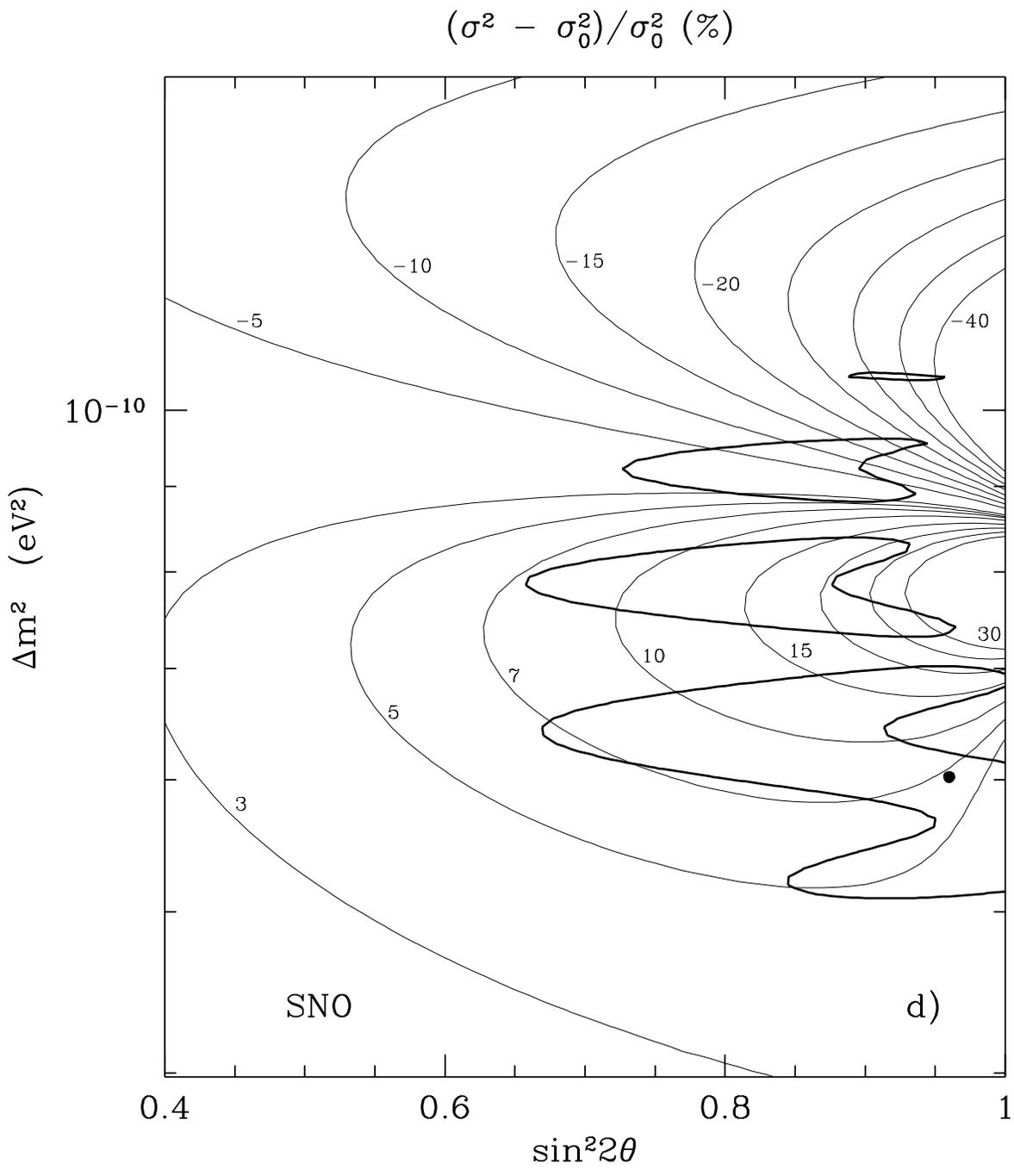
\end{document}